\documentclass[9pt,twocolumn,twoside]{opticajnl}
\journal{opticajournal} 

\usepackage{xcolor}
\usepackage{lineno}
\usepackage{xspace}
\usepackage{siunitx}
\usepackage{multirow}
\usepackage{graphicx}
\graphicspath{{figures/}, {images/}}

\newcommand{\algo}{ODkAnon\xspace}

\title{Protecting participants or population?  Comparison of $k$-anonymous Origin-Destination matrices}

\author[1]{Pietro Armenante}
\author[1]{Kai Huang}
\author[1]{Nikhil Jha}
\author[1,*]{Luca Vassio}

\affil[1]{Politecnico di Torino, Italy}
\affil[*]{luca.vassio@polito.it}

\setboolean{displaycopyright}{false} 

\begin{document}

\maketitle

\section{Introduction}
\label{sec:introduction}

Mobility data have become increasingly accessible due to the evolution of data collection processes and the diversification of their sources, such as the widespread adoption of GPS devices and mobile phones. 
Human mobility refers to the study of how people move, for instance, by characterizing patterns such as commuting to work, returning home, or using public transportation. A thorough understanding of these patterns is fundamental in several domains, including epidemic control and public transportation planning. 

Origin-Destination (OD) matrices are a core component of research on users' mobility and summarize how individuals move between geographical regions. 
OD-matrices describe the flows between origins \textit{o} and destinations \textit{d}. Although they represent a dramatic simplification compared to original individual movements from raw data, they are still a crucial indicator of mobility and expose privacy vulnerabilities. They are often harder to anonymise than regular relational data \cite{matet2023adaptative}.

This work is centered around the NetMob 2025 data challenge dataset \cite{netmob25}.\footnote{\url{https://netmob.org/www25/datachallenge}, accessed on \today} There are two added values in the provided data. Firstly, the data is extensive and contains a lot of socio-demographic information that can be used to create multiple OD matrices, based on the segments of the population. Secondly, and more peculiarly, a participant is not merely a record in the data, but a quantified statistically weighted proxy for a segment of the real population.  
This second aspect opens the door to a fundamental shift in the anonymization paradigm. 
A population-based view of privacy is central to our contribution. By adjusting our anonymization framework to account for representativeness, we are also protecting the inferred identity of the actual population, rather than survey participants alone.

The generation of privacy-preserving OD matrices has already been studied from various points of view. 
The works in~\cite{kim2024distod,matet2023adaptative} use local differential privacy to compute OD-matrices. Another work~\cite{boninsegna2025differentially} adapts the TopDown algorithm used in the US census~\cite{abowd20222020} to generate differentially private hierarchical OD data with high utility. Our approach builds upon the approach and insights from Matet et al. \cite{matet2023adaptative}, which propose an Adaptative Tree Generalisation (ATG) approach to anonymize OD matrices, optimizing the hierarchy generalization and comparing it with OIGH and Mondrian. Unlike these approaches, which primarily address the privacy aspects of the given datasets, we aim to contribute to the generation of privacy-preserving OD matrices enriched with socio-demographic segmentation that achieves $k$-anonymity on the actual population.

The challenge addressed in this work is to produce and compare OD matrices that are  $k$-anonymous for survey participants and for the whole population. Put briefly, $k$-anonymity~\cite{sweeney2002achieving}  is a privacy property of the data: a dataset is $k$-anonymous if every record cannot be distinguished among at least $k-1$ others---in our case, if there exist at least $k$ trips from the same origin to the same destination.

We compare the OD matrices obtained by several traditional methods of anonymization to reach $k$-anonymity by generalizing geographical areas. These include generalization over a hierarchy  (ATG and OIGH) and the classical Mondrian. To this established toolkit, we add a novel method, i.e., \algo, a greedy algorithm aiming at balancing speed and quality. 
This novel approach is benchmarked against three well-known generalisation algorithms, evaluating both individual-level and population-level privacy through weighted mobility data. Many utility metrics are considered in the comparison. This comprehensive analysis demonstrates how significant differences may exist when considering population-protecting for OD matrices anonymization, rather than survey participant-protecting ones. Moreover, we show how these differences can even be amplified across socio-demographic segments.

All the obtained results are reproducible using our open-source code available in a GitHub repository.\footnote{ \url{https://github.com/SmartData-Polito/ODkAnon}, accessed on \today.}

\section{Dataset and Methodology}

\subsection{Geographical area H3 hierarchy}
\label{sec:h3}

In this work, we use the well-known, Uber-developed H3 hierarchy.\footnote{\url{https://h3geo.org/docs/}, accessed on \today.} In H3, data points are bucketed into hexagons. 
Uber decided to create H3, combining the benefits of a hexagonal global grid system with a hierarchical indexing system.

\begin{figure}
    \centering
    \includegraphics[width=\linewidth]{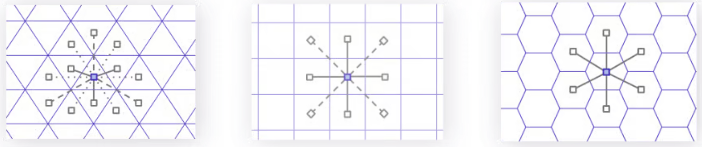}
    \caption{Triangular, square, and hexagon tiles. Distance between neighbors' centerpoints is highlighted, with hexagons neighbors' distance being constant (image from \url{https://www.uber.com/en-IT/blog/h3/}).}.
    \label{fig:importanceOfHexagons}
\end{figure}

Using a hexagon as the cell shape is critical for H3. As depicted in Figure \ref{fig:importanceOfHexagons}, hexagons have only one distance between a hexagon centerpoint and its neighbors' centerpoints, compared to two distances for squares or three distances for triangles. This property greatly simplifies performing analysis and smoothing over gradients.

\begin{figure}
\centering
\fbox{\includegraphics[width=0.8\linewidth]{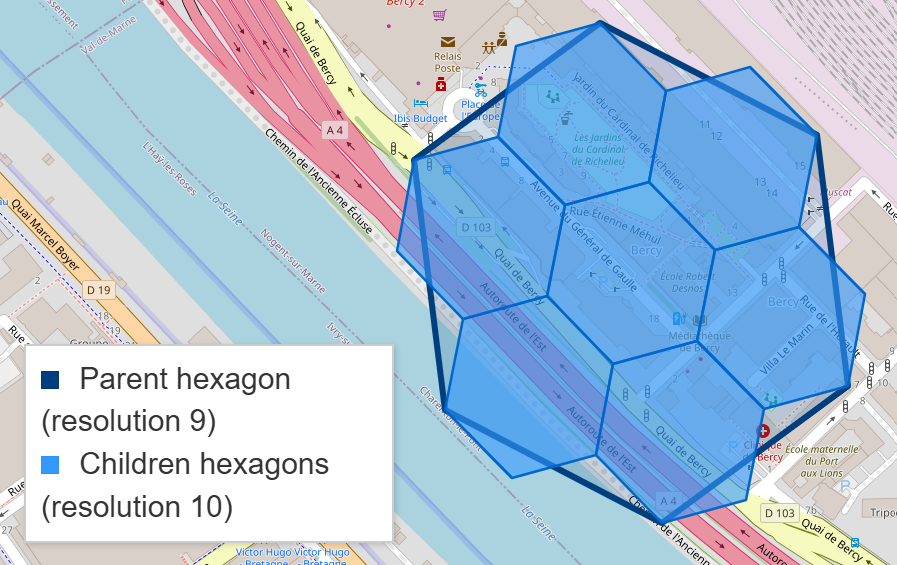}}
\caption{Example of a H3 hierarchy of hexagons in Paris.}
\label{fig:1}
\end{figure}

H3 supports sixteen resolutions---we sum them up in Table~\ref{tab:h3_index}. Each finer resolution has cells with one seventh the area of the coarser resolution. Hexagons cannot be perfectly subdivided into seven hexagons, so the finer cells are only approximately contained within a parent cell. We show an example of how the hierarchy is organized on a portion of the Paris map in Figure~\ref{fig:1}.

\begin{table}[h]
\centering
\caption{Total number of cells covering the Earth's surface and the corresponding area in $km^2$ for each level of the Uber H3 hierarchy. The levels highlighted in red are the ones taken into consideration for the NetMob 2025 Data Challenge. Resolution 4 captures the whole Île-de-France region.}
\label{tab:h3_index}
\small
\begin{tabular}{c|r|r}
\hline
\textbf{H3 index} & \textbf{Number of cells} & \textbf{Average hexagon area (km²)} \\
\hline
0 & 122 & 4357449.416078381 \\
1 & 842 & 609788.441794133 \\
2 & 5,882 & 86801.780398997 \\
3 & 41,162 & 12393.434655088 \\
\textcolor{red}{4} & \textcolor{red}{288,122} & \textcolor{red}{1770.347654491} \\
\textcolor{red}{5} & \textcolor{red}{2,016,842} & \textcolor{red}{252.903858182} \\
\textcolor{red}{6} & \textcolor{red}{14,117,882} & \textcolor{red}{36.129062164} \\
\textcolor{red}{7} & \textcolor{red}{98,825,162} & \textcolor{red}{5.161293360} \\
\textcolor{red}{8} & \textcolor{red}{691,776,122} & \textcolor{red}{0.737327598} \\
\textcolor{red}{9} & \textcolor{red}{4,842,432,842} & \textcolor{red}{0.105332513} \\
\textcolor{red}{10} & \textcolor{red}{33,897,029,882} & \textcolor{red}{0.015047502} \\
11 & 237,279,209,162 & 0.002149643 \\
12 & 1,660,954,464,122 & 0.000307092 \\
13 & 11,626,681,248,842 & 0.000043870 \\
14 & 81,386,768,741,882 & 0.000006267 \\
15 & 569,707,381,193,162 & 0.000000895 \\
\hline
\end{tabular}
\end{table}

\subsection{NetMob 2025 Dataset}
\label{sec:test}

The NetMob2025 challenge dataset \cite{netmob25} offers a multi-dimensional view of mobility behavior in the Île-de-France region over a continuous 7-day period. We select only participants with complete GNSS records. Then, we consider a trip completed after the absence of GNSS records for 3 minutes. Finally, we extract the origin and destination coordinates of each trip. 
These are anonymized, generalized to the center of the closest H3 hexagon of resolution~10.

Together with origins and destinations, we employ the weight feature assigned to every trip, describing how many people the record is representative of, and three socio-demographic features (age, sex, and profession) that will allow us to build segment-specific OD matrices. The main characteristics relevant to our work are reported in Table~\ref{tab:1}.


\begin{table}
\centering
\small
\caption{\bf Challenge filtered dataset relevant characteristics.}
\begin{tabular}{cc}
\hline
Participants with GNSS records & 3,320 \\
Represented population & 9,001,164 \\
Trips with GNSS points & 81,291 \\ 
H3 hexagons at resolution 10 & 44,011 \\
Relevant socio-demographic features & age, sex, profession  \\
\hline
\end{tabular}
  \label{tab:1}
\end{table}

The final dataset we employ is a merge of a dataset containing GNSS data and a dataset containing demographic data. Particularly, merging the two dataframes on the \texttt{person\_id}, it is possible to obtain a structure that contains both the GPS data and the demographic data. The dataset collects data obtained between October 2022 and May 2023 and focuses on residents aged 16 to 80 in Ile-de-France (3337 participants took part in the collection). The value that is more useful in the demographic data is the \texttt{WEIGHT\_INDIV}. Each participant is assigned a weight representing how many individuals in the Ile-de-France region share the same socio-demographic profile. This profile is defined by the cross-tabulation of several variables: department of residence (8 departments), age group (16–25,26–45,46–65,66–80), sex (male, female), socio-professional category (craftsmen, executives, intermediate professions, employees and workers, retirees and other inactives), number of cars in the household (0, 1, or 2+), household size (1, 2, 3, or 4+ people), and highest diploma obtained (lower or upper secondary, Bac+2, Bac+5 or doctorate). The final dataset will contain a person identifier, the GPS data (with the time) and the weight. 
The dataset initially contains 81,291 rows and 8 columns.

In a first iteration of our experiments, we will observe how the anonymization impacts the dataset as a whole. Then, we will study whether such an impact is uniform across different segments. To this end, we will consider the sex, age, and socio-professional category segmentations.

\subsection{Protecting the participants vs. protecting the population}

Common algorithms aim at obtaining a $k$-anonymous dataset, where every item---in this case, a trip---cannot be distinguished from at least $k$-1 other ones. However, each dataset only represents a subsample of the population, some segments of which may be over- or under-represented by the records in the dataset. The awareness of this data skewness could help re-identification attacks by an adversary. 
Therefore, risks are not only towards the participant from whom the data has been collected, but also towards the original population that has been sampled.


The weight attribute that the NetMob2025 challenge dataset offers hints at a solution in this direction, indicating the number of people in the Île-de-France area that each participant represents.  
By using this number to weight each entry in the dataset, we can tweak the algorithms to account for the population rather than the dataset, strengthening the privacy guarantees of the algorithms. Naively, the algorithms will see the same trip repeated as many times as the survey participant's population representativeness.
In our experiments, we will compare the two outcomes.

Notice that all the utility metrics we present in Section \ref{sec:metrics} can be computed either considering the survey participants (dataset records) or the underlying represented population, expressed by the weights. 

\subsection{Benchmark algorithms}

We compare our proposed \algo algorithm (detailed in Section \ref{sec:algo}) to several well-established algorithms in the field.

\begin{itemize}
\item Mondrian --- A partitioning algorithm~\cite{lefevre2006mondrian} which frames input data inside $k$-anonymous bounding hypercubes. We re-implemented the algorithm.

\item OIGH --- A heuristic, uniform tree generalization algorithm presented in~\cite{mahanan2021data}, which applies horizontal cuts within the hierarchies of origins and destinations. With this algorithm, all origins are generalized to the same level, which could be different from the destinations (and viceversa). We used the implementation in~\cite{matet2023adaptative} provided by the authors.

\item ATG --- Similarly to \algo, ATG is an adaptive generalization algorithm for OD matrices introduced in~\cite{matet2023adaptative}. We run the algorithm using the code provided by the authors of~\cite{matet2023adaptative}. We used the fastest version ATG--Soft to limit computational times. Still, we did not reach a solution for this algorithm in many scenarios (see Section  \ref{sec:results}).

\end{itemize}

Notice that we will use the same H3 hierarchy in \algo, ATG-Soft, and OIGH. Mondrian does not use any hierarchy.

ATG-Soft and Mondrian create \textit{non-homogeneous} geographical areas, i.e., origin areas choice depends on the given destination and vice versa. This means that the final OD matrix might comprise overlapping areas.
Our \algo algorithm and OIGH create \textit{homogeneous} areas, adding a constraint with respect to ATG-Soft and Mondrian.

Since the relative sparsity of OD matrices makes it very difficult to fully anonymize the data without an unacceptable amount of generalization, all the algorithms but Mondrian---\algo included---must consider the costs and opportunities of suppressing some data (i.e., trips) to obtain a useful, private aggregation.

\subsection{Performance indicators}
\label{sec:metrics}

The anonymization alters the values of the OD matrix, as well as the zone sizes, depending on the aggregation of the geographical areas in the hierarchy. We need to assess data utility and privacy preservation with respect to the original data. 

When applying anonymization techniques to protect data, it is essential to ensure that the process introduces only the minimum amount of generalization or perturbation required to satisfy the \textit{k}-anonymity constraint. While hiding sensitive data is a priority, the resulting dataset must remain useful for further analysis. Excessive perturbation may strengthen privacy protection, but it can also compromise data utility, making it difficult to extract meaningful insights. For this reason, it is crucial to compute some metrics that evaluate the quality and usability of data.

Privacy metrics validate the strength of the anonymization.  
We measure the minimum k-anonymity obtained for participants when protecting the population, and the other way around.

Data utility metrics measure how well the anonymized OD matrix preserves the original data characteristics, comparing the pre- and post-anonymization versions. The following are the metrics we used in this work.

All the metrics we presented can be computed either considering the dataset records or the underlying population, expressed by the weights: in every calculation, instead of counting for one, each row in the dataset counts for the number of people it represents---its weight. Each metric can be calculated in a survey participant-oriented or population-oriented way, independently of whether it was anonymized to protect the dataset or the population.

\subsubsection{Discernability Metric}

The first metric is one that attempts to capture in a straightforward way the desire to maintain discernibility between tuples as much as is allowed by a given setting of \textit{k} \cite{bayardo2005data}. This discernibility metric ($C_{DM}$) assigns a penalty to each tuple based on how many tuples in the transformed dataset are indistinguishable from it. If an unsuppressed tuple is part of an equivalence class $Eq$ of size $|Eq|$, then that tuple is assigned a penalty of $|Eq|$. If a tuple is suppressed, then it is assigned a penalty of $|D|$, the size of the input dataset: in this way, a suppressed tuple cannot be distinguished from any other tuple in the dataset, hence it needs a penalization larger than any non-suppressed tuple ($|D|\geq|Eq|$). The metric is defined mathematically as:
\begin{equation}
    C_{DM} = \sum_{ Eq s.t. |Eq| \geq k} |Eq|^2 + \sum_{Eq s.t. |Eq| < k}|D||Eq|
\end{equation}
In this expression, the sets \textit{Eq} refer to the equivalence classes of tuples in \textit{D} induced by the anonymization. The first sum computes penalties for each non-suppressed tuple, the second for suppressed tuples. Small values of this metric indicate small equivalence classes and no suppression, with resulting OD matrices retaining more utility and information. Notice that this metric is not normalized, and can assume very large values.

\subsubsection{Normalized Average Equivalence Class Size}

We use the normalized average equivalence class size metric ($C_{AVG}$) defined in \cite{lefevre2006mondrian}. The $C_{AVG}$ metric evaluates the average size of equivalence classes in a dataset after anonymization, normalized with respect to the anonymity parameter \textit{k}. Formally, it is defined as:
\begin{equation}
    C_{AVG} = \dfrac{\left(\dfrac{|D^+|}{total\_equiv\_classes}\right)}{k}
\end{equation}
$D^+$ is the set of non-suppressed records, leading to  $|D^+|\leq |D|$, and $total\_equiv\_classes$ is the number of equivalence classes that respects k-anonymity. 
This metric measures how much the average class size exceeds the minimum anonymity requirement. A value of $C_{AVG} = 1$ indicates that, on average, equivalence classes contain exactly \textit{k} records, meaning the dataset is minimally compliant with the \textit{k}-anonymity constraint. Higher values of $C_{AVG}$ suggest that equivalence classes are significantly larger than the threshold, which implies stronger anonymity but may also lead to greater information loss due to excessive generalization.

Notice that this metric does not penalize suppression and does not compare the obtained generalization with the original data.

\subsubsection{Mean Generalization Error}
The mean generalization error is defined by~\cite{matet2023adaptative} as follows:
\begin{equation}
    \bar{G} = \frac{1}{|D^+|}\sum_{Eq_{o\xrightarrow{}d} s.t. \left|Eq_{o\xrightarrow{}d}\right| \geq k}(|o| + |d|)\left|Eq_{o\xrightarrow{}d}\right| 
\end{equation}
where $D^+$ is the set of non-suppressed records and and $|D^+|$ is their total volume. We consider all non-suppressed equivalence classes $Eq_{o\xrightarrow{}d}$ that respect k-anonymity ($Eq_{o\xrightarrow{}d} \geq k$), which means only the ones belonging to $D^+$. 
$|o|$ and $|d|$ represent the number of original areas aggregated over the hierarchy, for origins and destinations, respectively.
This metric provides an estimation of the average information loss introduced during the generalization process. A larger value of \textit{$\bar{G}$} indicates that origins and/or destinations have been generalized into coarser spatial units, reducing the precision of the mobility representation. On the other hand, lower values of \textit{$\bar{G}$} reflect finer partitions, which preserve more spatial detail but may offer weaker privacy guarantees. 
Notice that this metric cannot be computed for algorithms that do not use a hierarchy.

\subsubsection{Reconstruction Loss}
This metric quantifies how much the generalized data deviates from the original data. The reconstruction loss \textit{E} is computed in this way \cite{matet2023adaptative}:
\begin{equation}
    E = \frac{1}{|D|} \sum_{o,d \in leaves(T)} ||\tilde{D}_{o \xrightarrow{} d}| - |D_{o \xrightarrow{} d}||
\end{equation}
where $T$ is the hierarchy tree, and we consider the finer-grained tiles in this hierarchy for the origins and destinations, i.e., the leaves of the tree.
We denote by $D_{o \xrightarrow{} d}$ the number of records in these original finer-grained tiles. 
Considering the leaves of the tree $T$ ensures that the loss is computed with respect to the original maximum granularity (level 10 of the H3 hierarchy), regardless of how much aggregation has been performed.
The anonymized coarser-grained equivalence classes 
$Eq_{o \xrightarrow{} d}$ that respect k-anonymity ($Eq_{o\xrightarrow{}d} \geq k$) should be mapped to the finer-grained ones. 
For every pair of origin cells $(o,d)$ at the maximum granularity (the leaves of the hierarchical structure $T$), we look up which generalised pair $(o',d')$ they belong to after anonymisation.
However, such information is now aggregated over multiple tiles, and we cannot infer the exact original values from the anonymized version. Then, the best we can do is to uniformly assign the anonymized records to the finer-grained tiles, proportionally to their size. We call this reconstructed finer-grained volume of trips $\tilde{D}_{o \xrightarrow{} d}$.
We normalise the metric by the total volume of the flows \textit{|D|} for readability. Notice that D also includes the flows that have been suppressed during the anonymisation, and the finer-grained tiles without any trips that generalize to tiles with at least a trip.
In a scenario where, during the anonymization process, new trips are not created, i.e., they do not increase but are possibly suppressed, the worst possible obtained value is 2. 

Notice that this metric can also not be computed for algorithms that do not use a hierarchy.

\section{\algo algorithm for homogenous k-anonymous OD matrices}
\label{sec:algo}

This section describes the ODkAnon algorithm for the anonymization of OD matrices, leveraging the Uber H3 geo-indexing library (we used H3 version 4.2.2). This is a novel algorithm, here presented and detailed for the first time, with open-source code available in our GitHub repository. 
In short, the algorithm 
iteratively selects and aggregates cells related to lower-density areas (considering both origins and destinations) until the $k$-anonymity condition is met.

This algorithm (as OIGH) creates \textit{homogeneous} geographical areas, i.e., origin areas choice does not depend on the given destination and vice-versa. This means that the final OD matrix will comprise non-overlapping areas. 
This is an additional constraint with respect to other techniques like Mondrian and ATG, for which the obtained anonymized hierarchy level for the destinations changes depending on the origin and viceversa (i.e., non-homogeneous generalization).

\algo applies an adaptive approach that dynamically decides, for each flow not meeting the \textit{k}-anonymity threshold, whether to generalize the origin or the destination hexagon. This decision is based on a greedy strategy that chooses the aggregation target minimizing the total trip count of candidate hexagons, thereby reducing spatial distortion while preserving data utility. To efficiently handle very large and sparse OD matrices, the algorithm leverages sparse matrix representations and precomputed hierarchical relationships within the H3 hexagonal index trees. This allows for rapid identification of sibling hexagons and their parents to speed up the generalization process and balance the matrix dimensions dynamically to avoid excessive loss of spatial resolution on either origin or destination side.

\subsection{Suppression algorithm}
Since hexagons may be very sparse or associated with very low counts (even when their siblings contain large volumes), they may need to be progressively generalized into larger and larger parent hexagons.

The algorithm first performs a pre-filtering step that identifies and removes OD pairs (origin–destination) that cannot achieve k-anonymity even after multiple levels of spatial generalization. The goal is to suppress those records that may lead to huge generalizations that make the data lose utility. 

The user can define a suppression budget as the maximum percentage of rows that one can remove, and a maximum number of generalization levels (\texttt{max generalization levels} L) to explore to understand if, within these levels, the node can be \textit{k}-anonymized. The algorithm---summarized in Algorithm~\ref{code:suppression_function}---works as follows.
\begin{itemize}
    \item \textbf{Generalization Mapping (row~\ref{code:gen})}. For each OD pair, the algorithm computes parent hexagons at progressively coarser resolutions (from 0---original hexagons---up to \texttt{max generalization levels} L).
    \item \textbf{Aggregated Counts (row~\ref{code:group})}. At each generalization level, the OD pairs are grouped by their generalized hexagons, and the aggregated counts are computed. If a group reaches or exceeds the threshold \textit{k}, the corresponding rows are marked as valid.
    \item \textbf{Detection of Problematic Rows (row~\ref{code:prob})}. Rows that do not reach the threshold $k$ at any generalization level are identified as problematic, meaning they cannot be made $k$-anonymous.
    \item \textbf{Suppression Strategy (row~\ref{code:suppress})}. If the number of problematic rows is within the suppression budget, all of them are removed. If not, the algorithm suppresses only the rows with the lowest counts, ensuring minimal information loss.
\end{itemize}
The function returns a filtered OD matrix where only valid (\textit{k}-anonymous) rows remain, along with the number of suppressed rows.

\begin{algorithm}[t]
\caption{Suppression algorithm.}
\label{code:suppression_function}
\begin{algorithmic}[1]
\Require OD matrix, threshold $k$, max generalization levels $L$, suppression budget $\beta$
\Ensure Filtered OD matrix

\State $n \gets |OD|$, $\text{max\_supp} \gets \lfloor n \cdot \beta \rfloor$
\State Initialize H3 mapping cache

\For{$\ell = 0$ \textbf{to} $L$} \label{code:gen}
    \For{each row in OD}
        \State $\text{start\_gen}_\ell \gets \text{generalize}(\text{start\_h3}, \ell)$ 
        \State $\text{end\_gen}_\ell \gets \text{generalize}(\text{end\_h3}, \ell)$
    \EndFor
\EndFor

\State $\text{valid\_pairs} \gets \emptyset$

\For{$\ell = 0$ \textbf{to} $L$} \label{code:group}
    \State Group by $(\text{start\_gen}_\ell, \text{end\_gen}_\ell)$
    \State Compute $\text{agg\_count}$ for each group
    \State $\text{valid\_pairs} \gets \text{valid\_pairs} \cup \{\text{rows with agg\_count} \geq k\}$
\EndFor

\State $\text{problematic} \gets \text{all\_rows} \setminus \text{valid\_pairs}$ \label{code:prob}

\If{$|\text{problematic}| \leq \text{max\_supp}$} \label{code:suppress}
    \State Suppress all problematic rows
\Else
    \State Sort problematic rows by increasing count
    \State Suppress the first $\text{max\_supp}$ rows
\EndIf

\Return OD matrix with valid rows only
\end{algorithmic}
\end{algorithm}

\subsection{Tree structure creation}
To handle the spatial generalization of OD matrices, we use two hierarchical H3 generalization trees: \texttt{tree\_start} and \texttt{tree\_end}. The main idea is to represent space not as a fixed grid, but as a tree structure, where each node corresponds to an H3 hexagon at a given resolution, and its children represent finer subdivisions.

The algorithm pseudo-code is reported in Algorithm~\ref{code:tree_function}
The inputs given to this part of the code are the OD matrix, the target resolution, which is the smallest H3 resolution considered (in this case is 10), and the column in the OD matrix to consider: first, the column containing the starting hexagon,s then the one considering the ending ones.

At this point, the algorithm follows these steps:
\begin{itemize}
    \item \textbf{Hexagons extraction}. Extract hexagons from the OD matrix dataset (row \ref{code:extract}).
    \item \textbf{Root identification}. Find the minimal optimal resolution (the biggest hexagon with count equal to 1): this hexagon will be the root of the tree (for cycle at row \ref{code:root}).
    \item \textbf{Hierarchy construction}. build the hierarchical paths from the minimum resolution up to the target resolution (for cycle at row \ref{code:construction}).
    \item \textbf{Node creation}. Create the nodes and establish the parent–child relationships between them (row \ref{code:node}).
    \item \textbf{Count population}. Populate the counts based on the OD matrix data: trip counts associated with each hexagon are inserted into the leaf nodes and then propagated upwards, so that each node represents the total number of trips across that area (for cycle at row \ref{code:count}).
\end{itemize}

\begin{algorithm}[t]
\caption{H3 Hierarchical Tree Construction.}
\label{code:tree_function}
\begin{algorithmic}[1]
\Require OD matrix, target resolution $R_{target}$, hex column $C$
\Ensure Optimized H3 hierarchical tree

\State $H \gets$ extract unique hexagons from column $C$ \label{code:extract}
\State $H_{coverage} \gets$ obtain full coverage at resolution $R_{target}$

\For{$r = 0$ \textbf{to} $R_{target}$} \label{code:root}
    \State $ancestors \gets \emptyset$
    \For{each $h \in H$}
        \State $ancestor \gets$ parent of $h$ at resolution $r$
        \State $ancestors \gets ancestors \cup \{ancestor\}$
    \EndFor
    \State $stats[r] \gets |ancestors|$
\EndFor

\State $R_{min} \gets \max\{r : stats[r] = 1\}$

\State $nodes \gets \emptyset$
\For{each $h \in H_{coverage}$} \label{code:construction}
    \State $path \gets$ path from $h$ to resolution $R_{min}$
    \For{each $p \in path$}
        \If{$p \notin nodes$}
            \State $nodes[p] \gets$ new H3 node \label{code:node}
        \EndIf
    \EndFor
    \State Establish parent-child relationships along $path$
\EndFor

\State $counts \gets$ group OD by column $C$ and sum counts
\For{each $(h, c) \in counts$} \label{code:count}
    \State $h_{target} \gets$ map $h$ to resolution $R_{target}$
    \State Propagate count $c$ from $h_{target}$ up to the root
\EndFor

\Return Hierarchical tree with aggregated counts
\end{algorithmic}
\end{algorithm}

\begin{algorithm}[h!]
\caption{Optimized OD Generalization (\texttt{OptimizedH3GeneralizedODMatrix})}
\label{code:optimized}
\begin{algorithmic}[1]

\Function{run\_optimized\_generalization}{$OD, tree_{start}, \linebreak tree_{end}, k$}
    \State Initialize sparse matrix with \Call{initialize\_optimized\_matrix}{}
    \State $step \gets 0$
    \While{minimum cell value $< k$}
        \State Select axis based on ratio balance (\emph{columns/rows alternation})
        \State $(group, parent, cost) \gets$ \Call{get\_best\_generalization\_fast}{axis}
        \If{no valid generalization found}
            \State Try alternative axis
            \If{still none}
                \State \textbf{break}
            \EndIf
        \EndIf
        \State \Call{apply\_sparse\_generalization}{group, parent, axis}
        \State $step \gets step + 1$
    \EndWhile
    \State \Return Final generalized sparse matrix
\EndFunction

\Function{initialize\_optimized\_matrix}{}
    \State Remove zero-count cells from $OD$
    \State Extract used start and end hexagons
    \State Map hexagons to target resolution using $tree_{start}, tree_{end}$
    \State Build sparse OD matrix $(rows, cols, counts)$
    \State Precompute sibling groups for both axes
    \State \Return sparse OD matrix
\EndFunction

\Function{get\_best\_generalization\_fast}{axis}
    \State $best \gets \infty$
    \For{each parent node in hierarchy (start or end)}
        \State $siblings \gets$ children of parent
        \State $present \gets$ siblings currently in matrix
        \If{$present \neq \emptyset$ and consistent with siblings}
            \State $cost \gets$ aggregated count of $present$
            \If{$cost < best$}
                \State update best group, parent, and cost
            \EndIf
        \EndIf
    \EndFor
    \State \Return $(group, parent, cost)$ if found, else None
\EndFunction

\Function{apply\_sparse\_generalization}{group, parent, axis}
    \If{axis = columns}
        \State Merge columns of $group$ into new column $parent$
        \State Update sparse matrix and start mappings
    \Else
        \State Merge rows of $group$ into new row $parent$
        \State Update sparse matrix and end mappings
    \EndIf
    \State Remove old sibling groups involving $group$
    \State Add new sibling group including $parent$ (if applicable)
\EndFunction

\end{algorithmic}
\end{algorithm}

\subsection{Generalization algorithm for \textit{k}-anonymity}
The class \texttt{OptimizedH3GeneralizedODMatrix} (pseudo-code in Algorithm~\ref{code:optimized}) aims to efficiently anonymize very large OD matrices. It integrates sparse data structures, hierarchical H3 trees, and dynamic balancing strategies.

The inputs to this core function are the OD matrix, the two trees created with the Algorithm \ref{code:tree_function}, and the parameter \textit{k} to satisfy \textit{k}-anonymity. The workflow of the algorithm is structured as follows:
\begin{itemize}
    \item \textbf{Initialization}. A sparse matrix representation (CSR/CSC format) is built, where rows correspond to destination hexagons and columns to origin hexagons. This drastically reduces memory usage compared to dense matrices.
    \item \textbf{Precomputation of Sibling Groups}. Using the tree structures, the algorithm precomputes sibling groups, i.e., sets of hexagons sharing the same parent. These groups represent the potential candidates for aggregation during the generalization process. Even single-child groups are included, ensuring that the algorithm can continue generalizing also when only one descendant is available.
    \item \textbf{Generalization Strategy}. At each iteration, the algorithm identifies the cell with the minimum count in the sparse OD matrix. If all cells are above the anonymity threshold \textit{k}, the process stops. Otherwise, a new generalization step is applied. To decide where to generalize, the algorithm uses a dynamic balancing strategy:
    \begin{itemize}
        \item It tracks the ratio between the number of origins and destinations.
        \item If this ratio deviates significantly (beyond $\pm3\%$) from its initial value, the algorithm forces generalization on the “dominant” axis (origins if too many columns, destinations if too many rows).
        \item Otherwise, it alternates between the two axes to maintain balance.
    \end{itemize}
    Within the chosen axis, the algorithm selects the best sibling group to aggregate, using a cost function based on the number of trips.
    \item \textbf{Application of Sparse Generalization}. The selected sibling group is merged into its parent node in the sparse matrix. The corresponding row(s) or column(s) are summed, and the matrix is updated while preserving efficiency.
    \item \textbf{Termination}. The process continues iteratively until every OD pair has at least k trips, or no further aggregation is possible.
\end{itemize}

\section{Results}
\label{sec:results}

Here we present the experiments we performed, which consider the following dimensions:

\begin{itemize}

    \item 4 different ways of segmenting the dataset: whole dataset, sex, age, and socio-professional category;

    \item 4 algorithms: \algo, ATG-Soft, OIGH, and Mondrian;

    \item 2 protection targets: participants, and population;
    
    \item 4 data utility metrics: Discernability Metric $C_{DM}$, Normalized Average Equivalence Class Size $C_{AVG}$,  Mean Generalization Error $\bar{G}$, and Reconstruction Loss $E$;

    \item 2 metrics evaluation computation: based on participants (trips), and based on population (trips multiplied by representativeness).
\end{itemize}

We focused only on the trips within the Île-de-France (starting and ending within the region). For the algorithms that allow suppression, we fixed a maximum threshold of 10\% of trip suppression.

For each run, we allow up to two hours of computation time. The runs that did not provide a result in time are reported as {N/A} in the tables.

\subsection{Results over the whole population}

We first run the \algo algorithm on the whole dataset. 
We start protecting the participants in the survey, and we set $k=10$ for obtaining a $k$-anonymous OD matrix.
For both the origins and the destination, we obtain 29 zones, merging the original thousands of resolution-10 hexagons according to the hierarchy.

When protecting the population, $k$ should be adapted, accounting for the representativeness of each participant. 
Given that a participant on average accounts for 2,674 people, we use $k=10\times 2,674$ in order to keep a fair comparison of the two approaches. 
We now obtain the same 29 destination zones, but more fine-grained 35 origin zones.

\begin{figure}
\centering
\fbox{\includegraphics[width=\linewidth]{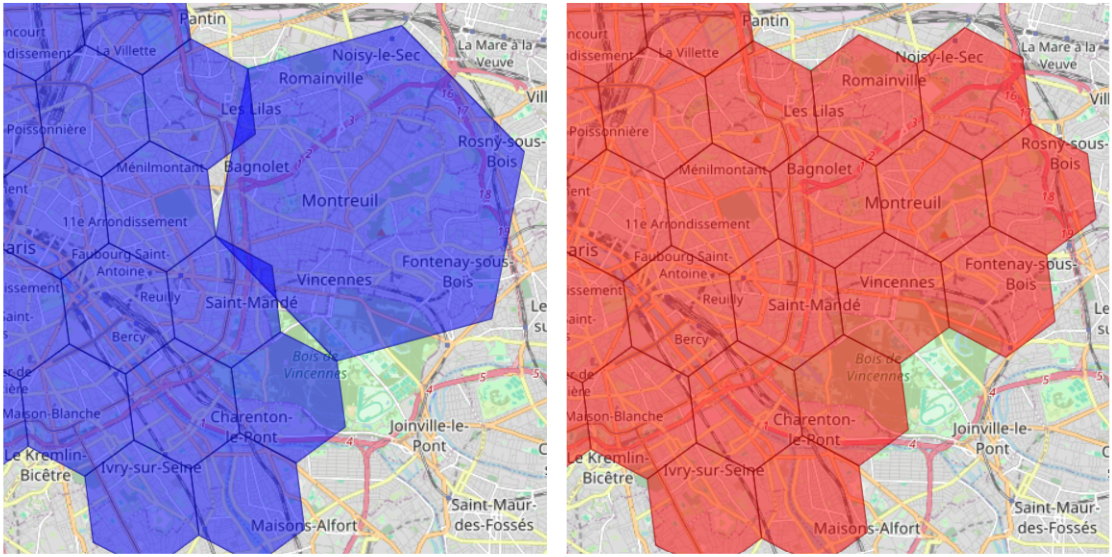}}
\caption{Detail over Paris of the origin generalization hexagons for the participant-protecting (left, blue hexagons) and the population-protecting (right, red hexagons) definitions.}
\label{fig:2}
\end{figure}

We show the comparison in the origin hexagons for the two approaches in Figure~\ref{fig:2}, with a zoom over the Paris region with observed differences. 
Protecting the population produces a different anonymization: in particular, 7 smaller hexagons (red tiles) are generalized to their parent node (blue tiles). 
Likely, these zones contain fewer trips from the participants---hence, they must be aggregated to satisfy $10$-anonymity---but said participants represent a sufficient amount of people, allowing to maintain a higher resolution when protecting the population.

Privacy metrics validate the strength of the anonymization.  
When generating the anonymous OD matrix for the participants, we also evaluate the impact on the population OD matrix, and vice versa. 
We measure the minimum k-anonymity obtained in such cases. We report the results of the privacy metric in Table~\ref{tab:whole-dataset-anonymity}.

The results show that protecting the participants leads to a population OD matrix that is no longer $k$-anonymous: 21 cells fall below the anonymity threshold, with a minimum value of 10,274 compared to the required 26,742. On the other hand, when the protection is applied to the population, the participants’ OD matrix does not reach the same level of anonymization $k=10$, as 13 cells fall below the threshold, and the minimal value is 4.
\begin{table}[t]
\centering
\caption{$k$-anonymity property computed in different scenarios.}
\begin{tabular}{rrrr}
\hline
 \multicolumn{2}{c}{\textbf{Participant-protecting}} & \multicolumn{2}{c}{\textbf{Population-protecting}}\\
 $k_{dataset}$ & $k_{population}$ &  $k_{dataset}$ & $k_{population}$ \\
\hline
10 &  10,274& 4 & 26,742\\
\hline
\end{tabular}
\label{tab:whole-dataset-anonymity}
\end{table}

We then compare the utility of data when applying different algorithms to the whole dataset. Table~\ref{tab:whole-dataset-dataset} shows the utility metrics in the survey participant-protecting scenario. In general, ATG-Soft and OIGH obtain relatively lower utility than \algo. Given that OIGH does not allow suppression, it over-generalizes sparse hexagons and thus loses more information. While ATG-Soft allows for suppression, its performance highly depends on the pre-definition of zones. As the authors of~\cite{matet2023adaptative} acknowledged, ATG-Soft needs proper tuning to achieve better performance. Mondrian does not consider H3 hexagons for hierarchy definition: it aggregates coordinates into rectangles. The more flexible generalization allows Mondrian to get the best performance on $C_{DM}$ and $C_{AVG}$, since it does not aggregate over hierarchy, we cannot derive $\bar{G}$ and $E$. While evaluating the results, it is important to keep in mind that non-homogeneity gives an advantage to ATG-Soft and Mondrian in terms of metrics, as it may happen that origin or destination overlap: while this offers greater flexibility to the algorithms, it may prevent real-world analysis of the results. For instance, if we want to understand the number of trips arriving to a zone, we might need to consider hexagons at different resolutions covering the zone.

We also evaluate the utility from the population's point of view. In general, the results are similar to calculating metrics on the participants, see Table~\ref{tab:whole-dataset-population} for more details. Notice that when computing $C_{DM}$ on population, the weights make it get much larger values, while the other metrics are normalized.

We report results on the population-protecting scenario in Table~\ref{tab:whole-population-dataset} and Table~\ref{tab:whole-population-population}. We were unable to get results from ATG-Soft and Mondrian within two hours of computation time. This is because now the number of trips is much larger (because the original trips are now multiplied by the representativeness), and the two algorithms scale poorly. In general, the utility metrics computed both on participants and population are on par with the participant-protecting scenario. In short, protecting participants or the population has little impact on data utility, but neither approach can guarantee the same level of privacy from the other perspective.



\begin{table}[ht]
\centering
\caption{Result comparison on the whole dataset, protecting the participants, calculating metrics on the participants. $\bar{G}$ and $E$ are not defined for Mondrian.}
\resizebox{\columnwidth}{!}{ 
\begin{tabular}{lrrrrr}
\hline
 & \multicolumn{1}{c}{$C_{DM}$ } & \multicolumn{1}{c}{$C_{AVG}$} & \multicolumn{1}{c}{$\bar{G}$} & \multicolumn{1}{c}{$E$} & Time (s) \\
\hline
\algo       & $5.4\times10^7$ & 13.2 &   601.8 &   1.91 & 26.4 \\
ATG-Soft    & $1.1\times10^8$ & 46.5 & 6,807.6 &  1.98  & 29.8 \\
OIGH        & $3.6\times10^7$ & 80.5 & 6,869.0 & 1.99 & 14.3 \\
Mondrian    & $4.0\times10^5$ &  1.3 &       - &       - &  7.2 \\
\hline
\end{tabular}
}
\label{tab:whole-dataset-dataset}
\end{table}

\begin{table}[ht]
\centering
\caption{Result comparison on the whole dataset, protecting the participants, calculating metrics on the population. $\bar{G}$ and $E$ are not defined for Mondrian.}
\resizebox{\columnwidth}{!}{ 
\begin{tabular}{lrrrrr}
\hline
 & \multicolumn{1}{c}{$C_{DM}$ } & \multicolumn{1}{c}{$C_{AVG}$} & \multicolumn{1}{c}{$\bar{G}$} & \multicolumn{1}{c}{$E$} & Time (s) \\
\hline
\algo       & $1.3\times10^{14}$ & 12.8 &   592.1 &   1.92 & 26.4 \\
ATG-Soft    & $2.2\times10^{14}$ & 44.2 & 6,824.5 &   1.88 & 29.8 \\
OIGH        & $2.4\times10^{14}$ & 77.6 & 6,867.3 & 1.99 & 14.3 \\
Mondrian    & $3.3\times10^{14}$ & 13.5 &       - &       - &  7.2 \\
\hline
\end{tabular}
}
\label{tab:whole-dataset-population}
\end{table}

\begin{table}
\centering
\caption{Result comparison on the whole dataset, protecting the population, calculating metrics on the participants. ATG--Soft and Mondrian did not provide a result within 2 hours of computation time.}
\resizebox{\columnwidth}{!}{ 
\begin{tabular}{lrrrrr}
\hline
 & \multicolumn{1}{c}{$C_{DM}$ } & \multicolumn{1}{c}{$C_{AVG}$} & \multicolumn{1}{c}{$\bar{G}$} & \multicolumn{1}{c}{$E$} & Time (s) \\
\hline
\algo       & $5.3\times10^7$ & 13.1 &   539.0 &   1.91 & 26.2 \\
ATG-Soft    &  \multicolumn{4}{c}{N/A} & >7,200.0  \\
OIGH        & $3.6\times10^7$ & 80.5 & 6,869.0 & 1.99 & 13.1 \\
Mondrian    & \multicolumn{2}{c}{N/A} & - & - & >7,200.0 \\
\hline
\end{tabular}
}
\label{tab:whole-population-dataset}
\end{table}

\begin{table}[ht]
\centering
\caption{Result comparison on the whole dataset, protecting the population, calculating metrics on the population. ATG--Soft and Mondrian did not provide a result within 2 hours of computation time.}
\resizebox{\columnwidth}{!}{ 
\begin{tabular}{lrrrrr}
\hline
 & \multicolumn{1}{c}{$C_{DM}$ } & \multicolumn{1}{c}{$C_{AVG}$} & \multicolumn{1}{c}{$\bar{G}$} & \multicolumn{1}{c}{$E$} & Time (s) \\
\hline
\algo       & $1.1\times10^{14}$ & 12.7 &   530.9 & 1.92 & 26.2 \\
ATG-Soft    &  \multicolumn{4}{c}{N/A} & >7,200.0  \\
OIGH        & $2.3\times10^{14}$ & 77.6 & 6,867.3 & 1.99 & 13.1 \\
Mondrian    & \multicolumn{2}{c}{N/A} & - & - & >7,200.0 \\
\hline
\end{tabular}
}
\label{tab:whole-population-population}
\end{table}

\subsection{Segmenting the population over sex}
Moreover, we have segmented the population taking into account three different attributes available in the NetMob dataset: sex, age, and profession. In particular, we divided the population by sex (men and women), by age groups (from 10 to 19 years old, from 20 to 29, from 30 to 39, from 40 to 49, from 50 to 59, from 60 to 99, and above 70), and by profession into eight categories. 

Results for the sex segmentation are reported in Tables~\ref{tab:sex-dataset-dataset}, \ref{tab:sex-dataset-population}, \ref{tab:sex-population-dataset}, and \ref{tab:sex-population-population}. In the next section, we will report the results for age and socio-professional category. Again, we imposed a two-hour deadline for the computation of every $k$-anonymized dataset. When protecting the population, ATG-Soft and Mondrian were not able to meet the time limit--hence, the metrics on these two algorithms were not evaluated.

When protecting the participants, anonymizing the male dataset produces a 2×5 matrix, whereas anonymizing the female dataset results in a 29×29 matrix. This indicates that protecting men is more challenging, as it requires very coarse hexagons, while for women the resulting hexagons remain much finer. Furthermore, when applying protection to the population, the difference becomes even more pronounced: the anonymized male dataset reduces to a 2×2 matrix.

We noticed how these differences are mainly observed when using \algo (see in particular Tables~\ref{tab:sex-dataset-dataset} and \ref{tab:sex-dataset-population}). Indeed, \algo is the only algorithm able to reach very high utility metrics for women. This observation raises the question of whether the difference lies in the data distribution or in the algorithms' choices, and will be evaluated in depth in future work.

\begin{table}[ht]
\centering
\caption{Result comparison segmenting on sex, protecting the participants, calculating metrics on the participants. $\bar{G}$ and $E$ are not defined for Mondrian.}
\begin{tabular}{lcrrrr}
\hline
& Sex & \multicolumn{1}{c}{$C_{DM}$ } & \multicolumn{1}{c}{$C_{AVG}$} & \multicolumn{1}{c}{$\bar{G}$} & \multicolumn{1}{c}{$E$} \\
\hline
\multirow{2}{*}{\algo}      & M & $4.6\times10^7$& 151.6 & 4,388.3 & 1.43\\
                            & F & $1.8\times10^7$ & 9.5 & 446.7 & 1.51\\ \hline
\multirow{2}{*}{ATG-Soft}   & M & $3.3\times10^7$ & 49.0 & 5,140.5 & 1.88\\
                            & F & $4.1\times10^7$ & 67.4 & 5,300.2 & 1.90\\ \hline
\multirow{2}{*}{OIGH}       & M & $8.0\times10^6$ & 37.9 & 4,843.3 & 1.99\\
                            & F & $1.0\times10^7$ & 42.6 & 5,033.7 & 1.99\\ \hline
\multirow{2}{*}{Mondrian}   & M &$ 1.7\times10^5$ & 1.3 & - & -\\
                            & F & $2.2\times10^5$ & 1.4 & - & -\\

\hline
\end{tabular}
\label{tab:sex-dataset-dataset}
\end{table}

\begin{table}[ht]
\centering
\caption{Result comparison segmenting on sex, protecting the participants, calculating metrics on the population. $\bar{G}$ and $E$ are not defined for Mondrian.}
\begin{tabular}{lcrrrr}
\hline
 & Sex & \multicolumn{1}{c}{$C_{DM}$ } & \multicolumn{1}{c}{$C_{AVG}$} & \multicolumn{1}{c}{$\bar{G}$} & \multicolumn{1}{c}{$E$} \\
\hline
\multirow{2}{*}{\algo}      & M & $2.1\times10^{14}$ & 141.4 & 4,377.5 & 1.87 \\
                            & F & $4.1\times10^{13}$ &   9.4 &   433.5 & 1.88 \\ \hline
\multirow{2}{*}{ATG-Soft}   & M & $1.0\times10^{14}$ &  45.3 & 5,180.0 & 1.88 \\
                            & F & $1.6\times10^{14}$ &  66.2 & 5,340.5 & 1.89 \\ \hline
\multirow{2}{*}{OIGH}       & M & $4.9\times10^{13}$ &  35.2 & 4,842.4 & 1.99 \\
                            & F & $7.2\times10^{13}$ &  42.3 & 5,031.6 & 1.99 \\ \hline
\multirow{2}{*}{Mondrian}   & M & $1.0\times10^{14}$ &  10.9 &       - &                 - \\
                            & F & $1.7\times10^{14}$ &  14.0 &       - &                 - \\
\hline
\end{tabular}
\label{tab:sex-dataset-population}
\end{table}

\begin{table}[ht]
\centering
\caption{Result comparison segmenting on sex, protecting the population, calculating metrics on the participants. ATG--Soft and Mondrian did not provide a result within 2 hours of computation time.}
\begin{tabular}{lcrrrr}
\hline
 & Sex & \multicolumn{1}{c}{$C_{DM}$ } & \multicolumn{1}{c}{$C_{AVG}$} & \multicolumn{1}{c}{$\bar{G}$} & \multicolumn{1}{c}{$E$} \\
\hline
\multirow{2}{*}{\algo}      & M & $1.3\times10^8$ & 302.8 & 6,726.8 & 1.86\\
                            & F & $1.8\times10^7$ &   9.9 &   450.3 &   1.87\\ \hline
\multirow{2}{*}{ATG-Soft}   & M & \multicolumn{4}{c}{\multirow{2}{*}{N/A}} \\
                            & F &  &  &  &\\ \hline
\multirow{2}{*}{OIGH}       & M & $8.0\times10^6$ &  37.9 & 4,843.3 &   1.99\\
                            & F & $1.0\times10^7$ &  42.6 & 5,033.7 &   1.99\\ \hline
\multirow{2}{*}{Mondrian}   & M & \multicolumn{2}{c}{\multirow{2}{*}{N/A}} & - & - \\
                            & F &  &  & - & - \\
\hline
\end{tabular}
\label{tab:sex-population-dataset}
\end{table}

\begin{table}[ht]
\centering
\caption{Result comparison segmenting on sex, protecting the population, calculating metrics on the population. ATG--Soft and Mondrian did not provide a result within 2 hours of computation time.}
\begin{tabular}{lcrrrr}
\hline
 & Sex & \multicolumn{1}{c}{$C_{DM}$ } & \multicolumn{1}{c}{$C_{AVG}$} & \multicolumn{1}{c}{$\bar{G}$} & \multicolumn{1}{c}{$E$} \\
\hline
\multirow{2}{*}{\algo}      & M & $7.5\times10^{14}$ & 292.5 & 6,724.4 & 1.92\\
                            & F & $4.1\times10^{13}$ &   9.9 &   437.2 & 1.89\\ \hline
\multirow{2}{*}{ATG-Soft}   & M & \multicolumn{4}{c}{\multirow{2}{*}{N/A}} \\
                            & F &  &  &  & \\ \hline
\multirow{2}{*}{OIGH}       & M & $4.9\times10^{13}$ & 35.2 & 4,842.4 & 1.99 \\
                            & F & $7.2\times10^{13}$ & 42.3 & 5,031.6 & 1.99 \\ \hline
\multirow{2}{*}{Mondrian}   & M & \multicolumn{2}{c}{\multirow{2}{*}{N/A}} & - & - \\
                            & F &  &  & - & - \\
\hline
\end{tabular}
\label{tab:sex-population-population}
\end{table}

\subsection{Other results}
In the following, we report the results for the age and the socio-professional category. For each segment, we keep $k=10$ for the $k$-anonymous OD matrix. Keeping $k = 10$ on much smaller datasets greatly reduces the data utility, and overgeneralizes the geographic area, leaving fewer than 5 origins/destinations.

In Tables~\ref{tab:age-dataset-dataset}, \ref{tab:age-dataset-population}, \ref{tab:age-population-dataset}, \ref{tab:age-population-population}, we show the results for the different combinations of protecting either the participants or the population, and calculating the metrics over either the participants or the population, segmenting the dataset according to participants' age. We show the same results, this time segmenting according to the socio-professional categories, in Tables~\ref{tab:cat-dataset-dataset}, \ref{tab:cat-dataset-population}, \ref{tab:cat-population-dataset}, \ref{tab:cat-population-population}.

Again, when protecting the population, ATG-Soft and Mondrian computation exceeded our two-hour computation limit.

\begin{table}[ht]
\centering
\caption{Result comparison segmenting on age, protecting the participants, calculating metrics on the participants. $\bar{G}$ and $E$ are not defined for Mondrian.}
\begin{tabular}{lcrrrr}
\hline
& Age & \multicolumn{1}{c}{$C_{DM}$ } & \multicolumn{1}{c}{$C_{AVG}$} & \multicolumn{1}{c}{$\bar{G}$} & \multicolumn{1}{c}{$E$} \\
\hline
\multirow{7}{*}{\rotatebox{90}\algo}        & [10-20] & $5.9\times10^5$ &  21.3 &   801.2 &   1.81\\
                                            & [20-30] & $3.1\times10^7$ & 156.9 & 4,154.3 & 1.85\\ 
                                            & [30-40] & $2.2\times10^7$ & 128.5 & 3,917.3 & 1.84\\ 
                                            & [40-50] & $8.6\times10^6$ &  66.8 & 2,481.9 &   1.84\\ 
                                            & [50-60] & $5.5\times10^6$ &  52.9 & 2,057.5 &   1.84\\ 
                                            & [60-70] & $6.7\times10^6$ &  69.3 & 2,583.5 & 1.83\\ 
                                            &     >70 & $9.5\times10^5$ &  26.2 & 1,082.6 &   1.83\\ \hline
\multirow{7}{*}{\rotatebox{90}{ATG-Soft}}   & [10-20] & $8.8\times10^5$ & 939.0 & 1,176.0 &   2.00\\
                                            & [20-30] & $1.1\times10^7$ & 137.6 & 3,703.9 & 1.99\\ 
                                            & [30-40] & $7.6\times10^6$ & 113.0 & 3,400.2 & 1.99\\ 
                                            & [40-50] & $1.0\times10^7$ &  65.4 & 3,427.5 & 1.89\\ 
                                            & [50-60] & $5.0\times10^6$ &  93.0 & 2,823.5 &   1.99\\ 
                                            & [60-70] & $2.9\times10^6$ &  46.3 & 2,213.8 &   1.54\\ 
                                            &     >70 & $1.3\times10^6$ & 115.1 & 1,467.0 & 1.99\\ \hline
\multirow{7}{*}{\rotatebox{90}{OIGH}}       & [10-20] & $2.2\times10^5$ &  18.7 &   743.6 &   1.99\\
                                            & [20-30] & $2.3\times10^6$ &  19.6 & 3,146.5 &   1.99\\ 
                                            & [30-40] & $1.4\times10^6$ &  16.1 & 2,808.3 &   1.99\\ 
                                            & [40-50] & $1.5\times10^6$ &  16.7 & 2,938.1 &   1.99\\ 
                                            & [50-60] & $1.0\times10^6$ &  13.2 & 2,349.6 &   1.99\\ 
                                            & [60-70] & $2.3\times10^6$ &  61.1 & 2,177.1 &   1.99\\ 
                                            &     >70 & $3.7\times10^5$ &  23.0 &   923.3 &   1.99\\ \hline
\multirow{7}{*}{\rotatebox{90}{Mondrian}}   & [10-20] & $1.3\times10^1$ &   1.4 &       - &      - \\
                                            & [20-30] & $9.5\times10^1$ &   1.3 &       - &      - \\ 
                                            & [30-40] & $6.7\times10^1$ &   1.1 &       - &      - \\ 
                                            & [40-50] & $7.0\times10^1$ &   1.1 &       - &      - \\ 
                                            & [50-60] & $7.9\times10^1$ &   1.6 &       - &      - \\ 
                                            & [60-70] & $3.7\times10^1$ &   1.2 &       - &      - \\ 
                                            &     >70 & $1.9\times10^1$ &   1.6 &       - &      - \\ \hline

\hline
\end{tabular}
\label{tab:age-dataset-dataset}
\end{table}

\begin{table}[ht]
\centering
\caption{Result comparison segmenting on age, protecting the participants, calculating metrics on the population. $\bar{G}$ and $E$ are not defined for Mondrian.}
\begin{tabular}{lcrrrr}
\hline
& Age & \multicolumn{1}{c}{$C_{DM}$ } & \multicolumn{1}{c}{$C_{AVG}$} & \multicolumn{1}{c}{$\bar{G}$} & \multicolumn{1}{c}{$E$}\\
\hline
\multirow{7}{*}{\rotatebox{90}{\algo}}      & [10-20] & $1.3\times10^{13}$ &  38.4 & 8,160.6 & 1.81 \\
                                            & [20-30] & $7.6\times10^{13}$ &  96.9 & 4,106.5 & 1.84 \\ 
                                            & [30-40] & $1.4\times10^{14}$ & 124.1 & 3,950.2 & 1.82 \\ 
                                            & [40-50] & $4.6\times10^{13}$ &  68.5 & 2,469.1 & 1.86 \\ 
                                            & [50-60] & $3.1\times10^{13}$ &  54.6 & 2,069.9 & 1.84 \\ 
                                            & [60-70] & $6.3\times10^{13}$ &  83.6 & 2,591.8 & 1.83 \\ 
                                            &     >70 & $6.4\times10^{12}$ &  27.8 & 1,054.1 & 1.78 \\ \hline
\multirow{7}{*}{\rotatebox{90}{ATG-Soft}}   & [10-20] & $2.0\times10^{13}$ & 168.4 & 1,176.0 & 2.00 \\
                                            & [20-30] & $3.0\times10^{13}$ &  85.6 & 3,704.0 & 1.99 \\ 
                                            & [30-40] & $5.0\times10^{13}$ & 110.2 & 3,385.1 & 1.99 \\ 
                                            & [40-50] & $5.4\times10^{13}$ &  65.7 & 3,451.3 & 1.88 \\ 
                                            & [50-60] & $4.0\times10^{13}$ &  95.8 & 2,829.9 & 1.99 \\ 
                                            & [60-70] & $2.3\times10^{13}$ &  55.7 & 2,221.9 & 1.57 \\ 
                                            &     >70 & $1.1\times10^{13}$ & 125.5 & 1,467.0 & 1.99 \\ \hline
\multirow{7}{*}{\rotatebox{90}{OIGH}}       & [10-20] & $5.3\times10^{12}$ &  33.8 &   744.7 & 1.99 \\
                                            & [20-30] & $6.4\times10^{12}$ &  12.2 & 3,146.5 & 1.99 \\ 
                                            & [30-40] & $9.6\times10^{12}$ &  15.7 & 2,804.5 & 1.99 \\ 
                                            & [40-50] & $1.1\times10^{13}$ &  17.0 & 2,937.8 & 1.99 \\ 
                                            & [50-60] & $8.1\times10^{12}$ &  13.6 & 2,349.5 & 1.99 \\ 
                                            & [60-70] & $2.5\times10^{13}$ &  73.6 & 2,175.1 & 1.99 \\ 
                                            &     >70 & $3.1\times10^{12}$ &  25.1 &   921.5 & 1.99 \\ \hline
\multirow{7}{*}{\rotatebox{90}{Mondrian}}   & [10-20] & $2.0\times10^{13}$ &  19.3 &       - &               - \\
                                            & [20-30] & $2.5\times10^{13}$ &   7.5 &       - &               - \\ 
                                            & [30-40] & $2.7\times10^{13}$ &   7.7 &       - &               - \\ 
                                            & [40-50] & $3.0\times10^{13}$ &   8.2 &       - &               - \\ 
                                            & [50-60] & $6.4\times10^{13}$ &  15.6 &       - &               - \\ 
                                            & [60-70] & $2.2\times10^{13}$ &   9.6 &       - &               - \\ 
                                            &     >70 & $1.4\times10^{13}$ &  14.2 &       - &               - \\ \hline
                
\hline
\end{tabular}
\label{tab:age-dataset-population}
\end{table}

\begin{table}[ht]
\centering
\caption{Result comparison segmenting on age, protecting the population, calculating metrics on the participants. ATG--Soft and Mondrian did not provide a result within 2 hours of computation time.}
\begin{tabular}{lcrrrr}
\hline
& Age & \multicolumn{1}{c}{$C_{DM}$ } & \multicolumn{1}{c}{$C_{AVG}$} & \multicolumn{1}{c}{$\bar{G}$} & \multicolumn{1}{c}{$E$}\\
\hline
\multirow{7}{*}{\rotatebox{90}{\algo}}      & [10-20] & $6.4\times10^5$ &  42.6 &   942.8 &   1.82 \\
                                            & [20-30] & $3.2\times10^7$ & 156.7 & 4,234.6 & 1.85 \\ 
                                            & [30-40] & $2.3\times10^7$ & 128.4 & 4,025.4 & 1.84 \\ 
                                            & [40-50] & $2.3\times10^7$ & 133.5 & 3,879.9 & 1.84 \\ 
                                            & [50-60] & $1.5\times10^7$ & 105.7 & 3,187.4 & 1.84 \\ 
                                            & [60-70] & $6.9\times10^6$ &  69.2 & 2,643.6 & 1.82 \\ 
                                            &     >70 & $9.6\times10^5$ &  26.1 & 1,105.8 &   1.83 \\ \hline
\multirow{7}{*}{\rotatebox{90}{ATG-Soft}}   & [10-20] & \multicolumn{4}{c}{\multirow{7}{*}{N/A}}    \\
                                            & [20-30] &                 &       &         &         \\ 
                                            & [30-40] &                 &       &         &         \\ 
                                            & [40-50] &                 &       &         &         \\ 
                                            & [50-60] &                 &       &         &         \\ 
                                            & [60-70] &                 &       &         &         \\ 
                                            &     >70 &                 &       &         &         \\ \hline
\multirow{7}{*}{\rotatebox{90}{OIGH}}       & [10-20] & $2.2\times10^5$ &  18.7 &   743.6 &   1.99 \\
                                            & [20-30] & $1.1\times10^7$ & 137.6 & 3,714.1 & 1.99 \\ 
                                            & [30-40] & $1.4\times10^6$ &  16.1 & 2,808.3 &   1.99 \\ 
                                            & [40-50] & $1.5\times10^6$ &  16.7 & 2,938.1 &   1.99 \\ 
                                            & [50-60] & $5.1\times10^6$ &  93.0 & 2,784.4 &   1.99 \\ 
                                            & [60-70] & $2.3\times10^6$ &  61.1 & 2,177.1 &   1.99 \\ 
                                            &     >70 & $3.7\times10^5$ &  23.0 &   923.3 &   1.99 \\ \hline
\multirow{7}{*}{\rotatebox{90}{Mondrian}}   & [10-20] & \multicolumn{2}{c}{\multirow{7}{*}{N/A}} & - &  -  \\
                                            & [20-30] &                 &       &       - &       - \\ 
                                            & [30-40] &                 &       &       - &       - \\ 
                                            & [40-50] &                 &       &       - &       - \\ 
                                            & [50-60] &                 &       &       - &       - \\ 
                                            & [60-70] &                 &       &       - &       - \\ 
                                            &     >70 &                 &       &       - &       - \\ \hline
\end{tabular}
\label{tab:age-population-dataset}
\end{table}

\begin{table}[ht]
\centering
\caption{Result comparison segmenting on age, protecting the population, calculating metrics on the population. ATG--Soft and Mondrian did not provide a result within 2 hours of computation time.}
\begin{tabular}{lcrrrr}
\hline
& Age & \multicolumn{1}{c}{$C_{DM}$ } & \multicolumn{1}{c}{$C_{AVG}$} & \multicolumn{1}{c}{$\bar{G}$} & \multicolumn{1}{c}{$E$}\\
\hline
\multirow{7}{*}{\rotatebox{90}{\algo}}      & [10-20] & $1.5\times10^{13}$ &  80.7 &   950.1 & 1.91 \\
                                            & [20-30] & $8.6\times10^{13}$ & 102.6 & 4,184.3 & 1.93 \\ 
                                            & [30-40] & $1.6\times10^{14}$ & 132.9 & 4,054.2 & 1.94 \\ 
                                            & [40-50] & $1.6\times10^{14}$ & 143.0 & 3,850.1 & 1.93 \\ 
                                            & [50-60] & $1.1\times10^{14}$ & 115.9 & 3,206.1 & 1.94 \\ 
                                            & [60-70] & $7.5\times10^{13}$ &  89.8 & 2,641.5 & 1.95 \\ 
                                            &     >70 & $7.9\times10^{12}$ &  30.4 & 1,083.1 & 1.94 \\ \hline
\multirow{7}{*}{\rotatebox{90}{ATG-Soft}}   & [10-20] &  \multicolumn{4}{c}{\multirow{7}{*}{N/A}}  \\
                                            & [20-30] &  &  &  &  \\ 
                                            & [30-40] &  &  &  &  \\ 
                                            & [40-50] &  &  &  &  \\ 
                                            & [50-60] &  &  &  &  \\ 
                                            & [60-70] &  &  &  &  \\ 
                                            &     >70 &  &  &  &  \\ \hline
\multirow{7}{*}{\rotatebox{90}{OIGH}}       & [10-20] & $5.3\times10^{12}$ &  33.8 &   744.7 & 1.99 \\
                                            & [20-30] & $3.1\times10^{13}$ &  85.6 & 3,714.0 & 1.99 \\ 
                                            & [30-40] & $9.6\times10^{12}$ &  15.7 & 2,804.5 & 1.99 \\ 
                                            & [40-50] & $1.1\times10^{12}$ &  17.0 & 2,937.8 & 1.99 \\ 
                                            & [50-60] & $4.1\times10^{13}$ &  95.8 & 2,791.7 & 1.99 \\ 
                                            & [60-70] & $2.5\times10^{13}$ &  73.6 & 2,175.1 & 1.99 \\ 
                                            &     >70 & $3.1\times10^{12}$ &  25.1 &   921.5 & 1.99 \\ \hline
\multirow{7}{*}{\rotatebox{90}{Mondrian}}   & [10-20] & \multicolumn{2}{c}{\multirow{7}{*}{N/A}} & - & - \\
                                            & [20-30] &  &  & - & - \\ 
                                            & [30-40] &  &  & - & - \\ 
                                            & [40-50] &  &  & - & - \\ 
                                            & [50-60] &  &  & - & - \\ 
                                            & [60-70] &  &  & - & - \\ 
                                            &     >70 &  &  & - & - \\ \hline

\hline
\end{tabular}
\label{tab:age-population-population}
\end{table}

\begin{table}[ht]
\centering
\caption{Result comparison segmenting on socio-professional category, protecting the participants, calculating metrics on the participants. $\bar{G}$ and $E$ are not defined for Mondrian.}
\begin{tabular}{lcrrrr}
\hline
& Cat. & \multicolumn{1}{c}{$C_{DM}$ } & \multicolumn{1}{c}{$C_{AVG}$} & \multicolumn{1}{c}{$\bar{G}$} & \multicolumn{1}{c}{$E$}\\
\hline
\multirow{8}{*}{\rotatebox{90}{\algo}}      & Cat. 1 & $9.1\times10^5$ &  24.6 & 1,131.1 &   1.82 \\
                                            & Cat. 2 & $2.3\times10^7$ & 110.0 & 3,666.6 & 1.86 \\ 
                                            & Cat. 3 & $1.2\times10^7$ &  95.8 & 3,034.3 & 1.84 \\ 
                                            & Cat. 4 & $8.8\times10^6$ &  82.7 & 2,486.3 & 1.84 \\ 
                                            & Cat. 5 & $5.2\times10^5$ &  38.1 &   940.5 &   1.81 \\ 
                                            & Cat. 6 & $6.3\times10^6$ &  68.5 & 2,451.1 & 1.83 \\ 
                                            & Cat. 7 & $4.7\times10^6$ &  48.1 & 1,842.0 &   1.84 \\ 
                                            & Cat. 8 & $5.9\times10^5$ &  17.5 &   789.1 &   1.81 \\ \hline
\multirow{8}{*}{\rotatebox{90}{ATG-Soft}}   & Cat. 1 & $1.1\times10^6$ & 108.5 & 1,428.0 &   2.00 \\
                                            & Cat. 2 & $2.2\times10^7$ &  62.0 & 4,545.4 & 1.97 \\ 
                                            & Cat. 3 & $4.1\times10^6$ &  84.2 & 2,655.1 &   1.82 \\ 
                                            & Cat. 4 & $4.3\times10^6$ &  45.3 & 2,230.0 &   1.88 \\ 
                                            & Cat. 5 & $7.0\times10^5$ &  84.0 & 1,142.0 &   2.00 \\ 
                                            & Cat. 6 & $2.4\times10^6$ &  60.3 & 2,175.9 &   1.99 \\ 
                                            & Cat. 7 & $4.3\times10^6$ &  84.4 & 2,535.3 &   1.99 \\ 
                                            & Cat. 8 & $2.3\times10^6$ & 154.4 & 2,001.0 & 2.00 \\ \hline
\multirow{8}{*}{\rotatebox{90}{OIGH}}       & Cat. 1 & $3.4\times10^5$ &  21.7 &   909.4 &   1.99 \\
                                            & Cat. 2 & $4.3\times10^6$ &  27.5 & 4,087.0 &   1.99 \\ 
                                            & Cat. 3 & $8.7\times10^5$ &  12.0 & 2,222.5 &   1.99 \\ 
                                            & Cat. 4 & $6.5\times10^5$ &  10.3 & 1,896.8 &   1.99 \\ 
                                            & Cat. 5 & $1.6\times10^5$ &  16.8 &   712.5 &   1.99 \\ 
                                            & Cat. 6 & $2.4\times10^6$ &  60.3 & 2,132.0 &   1.99 \\ 
                                            & Cat. 7 & $4.3\times10^6$ &  84.4 & 2,566.6 &   1.99 \\ 
                                            & Cat. 8 & $5.4\times10^5$ &  30.8 & 1,215.0 &   1.99 \\ \hline
\multirow{8}{*}{\rotatebox{90}{Mondrian}}   & Cat. 1 & $1.9\times10^1$ & 1.6 & - & - \\
                                            & Cat. 2 & $1.6\times10^2$ & 1.4 & - & - \\ 
                                            & Cat. 3 & $7.1\times10^1$ & 1.6 & - & - \\ 
                                            & Cat. 4 & $5.3\times10^1$ & 1.4 & - & - \\ 
                                            & Cat. 5 & $1.1\times10^1$ & 1.3 & - & - \\ 
                                            & Cat. 6 & $3.7\times10^1$ & 1.2 & - & - \\ 
                                            & Cat. 7 & $7.0\times10^1$ & 1.6 & - & - \\ 
                                            & Cat. 8 & $1.9\times10^1$ & 1.2 & - & - \\ \hline
\end{tabular}
\label{tab:cat-dataset-dataset}
\end{table}

\begin{table}[ht]
\centering
\caption{Result comparison segmenting on socio-professional category, protecting the participants, calculating metrics on the population. $\bar{G}$ and $E$ are not defined for Mondrian.}
\begin{tabular}{lcrrrr}
\hline
& Cat. & \multicolumn{1}{c}{$C_{DM}$ } & \multicolumn{1}{c}{$C_{AVG}$} & \multicolumn{1}{c}{$\bar{G}$} & \multicolumn{1}{c}{$E$}\\
\hline
\multirow{8}{*}{\rotatebox{90}{\algo}}      & Cat. 1 & $3.8\times10^{12}$ &  19.9 & 1,138.3 & 1.80 \\
                                            & Cat. 2 & $6.5\times10^{13}$ &  77.6 & 3,666.1 & 1.87 \\ 
                                            & Cat. 3 & $1.0\times10^{14}$ & 108.8 & 3,018.6 & 1.83 \\ 
                                            & Cat. 4 & $1.2\times10^{14}$ & 119.0 & 2,549.0 & 1.85 \\ 
                                            & Cat. 5 & $6.6\times10^{12}$ &  53.2 &   940.3 & 1.84 \\ 
                                            & Cat. 6 & $5.2\times10^{13}$ &  78.7 & 2,410.3 & 1.81 \\ 
                                            & Cat. 7 & $1.8\times10^{13}$ &  41.9 & 1,833.4 & 1.85 \\ 
                                            & Cat. 8 & $2.0\times10^{12}$ &  13.2 &   829.6 & 1.79 \\ \hline
\multirow{8}{*}{\rotatebox{90}{ATG-Soft}}   & Cat. 1 & $5.0\times10^{12}$ &  88.7 & 1,428.0 & 1.99 \\
                                            & Cat. 2 & $5.0\times10^{13}$ &  43.3 & 4,591.9 & 1.97 \\ 
                                            & Cat. 3 & $3.6\times10^{13}$ &  96.0 & 2,643.8 & 1.83 \\ 
                                            & Cat. 4 & $4.4\times10^{13}$ &  66.8 & 2,236.7 & 1.90 \\ 
                                            & Cat. 5 & $9.5\times10^{12}$ & 115.6 & 1,142.0 & 1.99 \\ 
                                            & Cat. 6 & $2.3\times10^{13}$ &  70.2 & 2,165.1 & 1.99 \\ 
                                            & Cat. 7 & $2.2\times10^{13}$ &  73.1 & 2,351.1 & 1.99 \\ 
                                            & Cat. 8 & $1.0\times10^{13}$ & 118.5 & 2,001.0 & 2.00 \\ \hline
\multirow{8}{*}{\rotatebox{90}{OIGH}}       & Cat. 1 & $1.5\times10^{12}$ &  17.7 &   898.4 & 1.99 \\
                                            & Cat. 2 & $1.5\times10^{13}$ &  19.3 & 4,089.8 & 1.99 \\ 
                                            & Cat. 3 & $8.0\times10^{12}$ &  13.7 & 2,220.0 & 1.99 \\ 
                                            & Cat. 4 & $1.0\times10^{13}$ &  14.8 & 1,895.7 & 1.99 \\ 
                                            & Cat. 5 & $2.4\times10^{12}$ &  23.1 &   715.6 & 1.99 \\ 
                                            & Cat. 6 & $2.3\times10^{13}$ &  70.2 & 2,121.4 & 1.99 \\ 
                                            & Cat. 7 & $2.2\times10^{13}$ &  73.1 & 2,561.2 & 1.99 \\ 
                                            & Cat. 8 & $2.7\times10^{12}$ &  23.7 & 1,223.9 & 1.99 \\ \hline
\multirow{8}{*}{\rotatebox{90}{Mondrian}}   & Cat. 1 & $7.3\times10^{12}$ &  10.7 &       - &               - \\
                                            & Cat. 2 & $7.9\times10^{13}$ &  10.9 &       - &               - \\ 
                                            & Cat. 3 & $5.7\times10^{13}$ &  16.2 &       - &               - \\ 
                                            & Cat. 4 & $5.7\times10^{13}$ &  15.3 &       - &               - \\ 
                                            & Cat. 5 & $7.9\times10^{12}$ &  11.2 &       - &               - \\ 
                                            & Cat. 6 & $2.1\times10^{13}$ &   9.4 &       - &               - \\ 
                                            & Cat. 7 & $4.4\times10^{13}$ &  13.7 &       - &               - \\ 
                                            & Cat. 8 & $3.3\times10^{12}$ &   4.9 &       - &               - \\ \hline
\end{tabular}
\label{tab:cat-dataset-population}
\end{table}

\begin{table}[ht]
\centering
\caption{Result comparison segmenting on socio-professional category, protecting the population, calculating metrics on the participants.ATG--Soft and Mondrian did not provide a result within 2 hours of computation time.}
\begin{tabular}{lcrrrr}
\hline
& Cat. & \multicolumn{1}{c}{$C_{DM}$ } & \multicolumn{1}{c}{$C_{AVG}$} & \multicolumn{1}{c}{$\bar{G}$} & \multicolumn{1}{c}{$E$}\\
\hline
\multirow{8}{*}{\rotatebox{90}{\algo}}      & Cat. 1 & $9.4\times10^5$ &  24.7 & 1,149.9 &   1.82 \\
                                            & Cat. 2 & $6.6\times10^7$ & 219.9 & 5,689.6 & 1.85 \\ 
                                            & Cat. 3 & $1.2\times10^7$ &  95.7 & 3,072.9 & 1.84 \\ 
                                            & Cat. 4 & $9.1\times10^6$ &  82.6 & 2,548.2 & 1.84 \\ 
                                            & Cat. 5 & $5.2\times10^5$ &  19.0 &   868.9 &   1.82 \\ 
                                            & Cat. 6 & $2.6\times10^6$ &  34.2 & 1,648.6 &   1.83 \\ 
                                            & Cat. 7 & $1.2\times10^7$ &  96.1 & 2,861.5 & 1.84 \\ 
                                            & Cat. 8 & $1.4\times10^6$ &  35.0 & 1,234.7 &   1.81 \\ \hline
\multirow{8}{*}{\rotatebox{90}{ATG-Soft}}   & Cat. 1 & \multicolumn{4}{c}{\multirow{8}{*}{N/A}}  \\
                                            & Cat. 2 &                 &       &         &         \\ 
                                            & Cat. 3 &                 &       &         &         \\ 
                                            & Cat. 4 &                 &       &         &         \\ 
                                            & Cat. 5 &                 &       &         &         \\ 
                                            & Cat. 6 &                 &       &         &         \\ 
                                            & Cat. 7 &                 &       &         &         \\ 
                                            & Cat. 8 &                 &       &         &         \\ \hline
\multirow{8}{*}{\rotatebox{90}{OIGH}}       & Cat. 1 & $3.4\times10^5$ &  21.7 &   909.4 &   1.99 \\
                                            & Cat. 2 & $4.3\times10^6$ &  27.5 & 4,087.0 &   1.99 \\ 
                                            & Cat. 3 & $8.7\times10^5$ &  12.0 & 2,222.5 &   1.99 \\ 
                                            & Cat. 4 & $3.1\times10^6$ &  75.5 & 2,234.4 &   1.99 \\ 
                                            & Cat. 5 & $1.6\times10^5$ &  16.8 &   712.5 &   1.99 \\ 
                                            & Cat. 6 & $2.4\times10^6$ &  60.3 & 2,132.0 &   1.99 \\ 
                                            & Cat. 7 & $4.3\times10^6$ &  84.4 & 2,566.6 &   1.99 \\ 
                                            & Cat. 8 & $5.4\times10^5$ &  30.8 & 1,215.0 &   1.99 \\ \hline
\multirow{8}{*}{\rotatebox{90}{Mondrian}}   & Cat. 1 & \multicolumn{2}{c}{\multirow{8}{*}{N/A}} & - & - \\
                                            & Cat. 2 &                 &       &      -  &       - \\ 
                                            & Cat. 3 &                 &       &      -  &       - \\ 
                                            & Cat. 4 &                 &       &      -  &       - \\ 
                                            & Cat. 5 &                 &       &      -  &       - \\ 
                                            & Cat. 6 &                 &       &      -  &       - \\ 
                                            & Cat. 7 &                 &       &      -  &       - \\ 
                                            & Cat. 8 &                 &       &      -  &       - \\ \hline
\end{tabular}-
\label{tab:cat-population-dataset}
\end{table}

\begin{table}[ht]
\centering
\caption{Result comparison segmenting on socio-professional category, protecting the population, calculating metrics on the population. ATG--Soft and Mondrian did not provide a result within 2 hours of computation time. }
\begin{tabular}{lcrrrr}
\hline
& Cat. & \multicolumn{1}{c}{$C_{DM}$ } & \multicolumn{1}{c}{$C_{AVG}$} & \multicolumn{1}{c}{$\bar{G}$} & \multicolumn{1}{c}{$E$}\\
\hline
\multirow{8}{*}{\rotatebox{90}{\algo}}      & Cat. 1 & $4.5\times10^{12}$ &  21.5 & 1,146.9 & 1.94 \\
                                            & Cat. 2 & $2.2\times10^{14}$ & 161.1 & 5,647.8 & 1.92 \\ 
                                            & Cat. 3 & $1.1\times10^{14}$ & 115.6 & 3,055.4 & 1.93 \\ 
                                            & Cat. 4 & $1.4\times10^{14}$ & 126.1 & 2,599.8 & 1.94 \\ 
                                            & Cat. 5 & $7.2\times10^{12}$ &  28.3 &   858.2 & 1.95 \\ 
                                            & Cat. 6 & $2.1\times10^{13}$ &  42.6 & 1,618.6 & 1.95 \\ 
                                            & Cat. 7 & $6.6\times10^{13}$ &  88.9 & 2,848.1 & 1.95 \\ 
                                            & Cat. 8 & $6.7\times10^{12}$ &  28.9 & 1,286.4 & 1.94 \\ \hline
\multirow{8}{*}{\rotatebox{90}{ATG-Soft}}   & Cat. 1 & \multicolumn{4}{c}{\multirow{8}{*}{N/A}}  \\
                                            & Cat. 2 &                    &       &         &                   \\ 
                                            & Cat. 3 &                    &       &         &                   \\ 
                                            & Cat. 4 &                    &       &         &                   \\ 
                                            & Cat. 5 &                    &       &         &                   \\ 
                                            & Cat. 6 &                    &       &         &                   \\ 
                                            & Cat. 7 &                    &       &         &                   \\ 
                                            & Cat. 8 &                    &       &         &                   \\ \hline
\multirow{8}{*}{\rotatebox{90}{OIGH}}       & Cat. 1 & $1.5\times10^{12}$ &  17.7 &   898.4 & 1.99 \\
                                            & Cat. 2 & $1.5\times10^{13}$ &  19.3 & 4,089.8 & 1.99 \\ 
                                            & Cat. 3 & $8.0\times10^{12}$ &  13.7 & 2,220.0 & 1.99 \\ 
                                            & Cat. 4 & $4.7\times10^{13}$ & 104.1 & 2,230.3 & 1.99 \\ 
                                            & Cat. 5 & $2.4\times10^{12}$ &  23.1 &   715.6 & 1.99 \\ 
                                            & Cat. 6 & $2.3\times10^{13}$ &  70.2 & 2,121.4 & 1.99 \\ 
                                            & Cat. 7 & $2.2\times10^{13}$ &  73.1 & 2,561.2 & 1.99 \\ 
                                            & Cat. 8 & $2.7\times10^{12}$ &  23.7 & 1,223.9 & 1.99 \\ \hline
\multirow{8}{*}{\rotatebox{90}{Mondrian}}   & Cat. 1 & \multicolumn{2}{c}{\multirow{8}{*}{N/A}} & - & - \\
                                            & Cat. 2 &                    &       &       - &                 - \\ 
                                            & Cat. 3 &                    &       &       - &                 - \\ 
                                            & Cat. 4 &                    &       &       - &                 - \\ 
                                            & Cat. 5 &                    &       &       - &                 - \\ 
                                            & Cat. 6 &                    &       &       - &                 - \\ 
                                            & Cat. 7 &                    &       &       - &                 - \\ 
                                            & Cat. 8 &                    &       &       - &                 - \\ \hline
\end{tabular}
\label{tab:cat-population-population}
\end{table}
\section{Conclusion and future work}

Unlike previous approaches, which primarily address the privacy aspects of a given dataset, we compared the generation of privacy-preserving OD matrices enriched with socio-demographic segmentation that achieves $k$-anonymity on the actual population.

We also proposed a new algorithm (\algo) that produces a k-anonymous homogeneous OD-matrix over a hierarchy. This algorithm proved better utility for most metrics than the only other homogeneous OIGH algorithm.

Our results showed that significant differences exist when protecting the population while creating anonymous OD matrices, rather than protecting survey participants. Moreover, we notice that protecting one of the two does not guarantee protection for the other: this seems to suggest that both definitions of privacy should be used together to offer a better understanding of the privacy guarantees offered by an algorithm. Moreover, we showed how these differences can even be amplified across socio-demographic segments. Some segments (such as men with respect to women) appear much more difficult to anonymize with respect to others, irrespective of their size in terms of trips.

For future work, the approach presented here should be replicated on other datasets related to human mobility. In particular, we plan to use a dataset describing a complete year of the trajectories for all the 442 taxis running in the city of Porto, Portugal, for a total of 1,710,670 trips\footnote{ \url{https://www.kaggle.com/datasets/crailtap/taxi-trajectory?resource=download}, accessed on \today.}. 
Our open-source code also allows other researchers and practitioners to easily apply the techniques presented here to their own datasets and case studies.

Finally, in this work, we observed the results of the combination of some algorithms on different segments of the dataset, and detected when reaching anonymity is more challenging. Future work might investigate theoretical limits intrinsic to the dataset and its distribution over space, irrespective of the used algorithm.






\begin{backmatter}

\bmsection{Acknowledgment} This work has been funded by the project “National Center for HPC, Big Data and Quantum Computing,” Grant No. CN00000013 (Bando M42C–Investimento 1.4–Avviso Centri Nazionali–D.D. No. 3138 of 16.12.2021, funded with MUR Decree No. 1031 of 17.06.2022), and by SmartData@Polito research center.





\bmsection{Disclosures} The authors declare no conflicts of interest.

\end{backmatter}



\bibliography{biblio}

\end{document}